\documentstyle[twocolumn,aps,epsfig]{revtex}

\begin{document}
\author{C. Figueira de Morisson Faria$^1$ and M. L. Du$^{1,2}$}
\address{$^1$Max Planck Institut f\"{u}r Physik komplexer Systeme,
N\"{o}thnitzer Str. 38, D-01187 Dresden, Germany\\
$^2$Institute of Theoretical Physics, P.O.Box 2735, Beijing, China}
\title{Enhancement of bichromatic high-harmonic generation with a high-frequency
field}
\date{\today}
\maketitle

\begin{abstract}
Using a high-frequency field superposed to a linearly polarized bichromatic
laser field composed by a wave with frequency $\omega $ and a wave with
frequency $2\omega $, we show it is possible to enhance the intensity of a
group of high harmonics in orders of magnitude. These harmonics have
frequencies about 30\% higher than the monochromatic-cutoff frequency, and,
within the three-step-model framework, correspond to a set of electron
trajectories for which tunneling ionization is strongly suppressed.
Particular features in the observed enhancement suggest that the
high-frequency field provides an additional mechanism for the electron to
reach the continuum. This interpretation is supported by a time-frequency
analysis of the harmonic yield. The additional high frequency field permits
the control of this group of harmonics leaving all other sets of harmonics
practically unchanged, which is an advantage over schemes involving only
bichromatic fields.
\end{abstract}

\vspace{0.2cm} 
\noindent 32.80.Rm, 32.80.Qk, 42.65.Ky, 42.50.Hz \vspace{0.3cm}

Within the last few years, the perspective of obtaining efficient laser
sources in the extreme ultraviolet (XUV) regime has led to the proposal of
several schemes for controlling the harmonic spectrum of an atom subject to
a strong laser field ($I\sim 10^{14}{\rm W/cm^{2}}$), using for instance
additional static fields \cite{static}, ultrashort pulses \cite{ushort},
bichromatic driving fields \cite
{bichro,cfbichro1,cfbichro2,doublemil,expebic,andiel}, or additional
confining potentials \cite{trap}. These schemes are based on the
``three-step model'' \cite{tstep}, which describes very well the spectral
features observed in experiments. These features are the ``plateau'', where
harmonics of roughly the same intensities exist, and the ``cutoff'', where
the harmonic signal suddenly decreases in orders of magnitude. According to
this model, the generation of high-order harmonics in strong laser fields
corresponds to a dynamical process in which an electron leaves an atom at an
instant $t_{0}$ via tunneling ionization, this electron in the continuum is
then accelerated by the driving fields, and, if it comes back and recombines
with the parent ion at a time $t_{1}$, a high energy photon will be
generated. The harmonic energy is given by $\Omega =|\varepsilon _{0}|+E_{%
{\rm kin}}(t_{0},t_{1}),$ with $E_{{\rm kin}}(t_{0},t_{1})$ and $%
|\varepsilon _{0}|$ being, respectively, the kinetic energy of the electron
upon return and the ionization potential of its parent ion. The cutoff
corresponds to the maximum electron kinetic energy. By manipulating the
different steps of the dynamical process, one can in principle control
high-harmonic generation.

If the purpose is to increase the cutoff energy, very efficient schemes
exist. They involve either additional static fields \cite{static}, or atoms
placed in confining potentials \cite{trap}. In both cases one introduces an
additional force which modifies the propagation of the electron after
injection, resulting in an increased cutoff energy. However, such schemes
may require extremely high static fields \cite{footnote1}, or appropriate
solid-state materials, whose existence is still under investigation \cite
{footnote2}.

Another scheme for manipulating high harmonics involves bichromatic driving
fields \cite{bichro,cfbichro1,cfbichro2,doublemil,expebic,andiel}. By varying the
frequencies, the relative phase, and the intensities of the two driving
waves, one can modify the propagation of the electron in the continuum, and
even the ``first step'', i.e., the tunneling ionization. In contrast to the
previously discussed schemes, high-harmonic generation with bichromatic
driving fields is already experimentally feasible \cite{expebic,andiel}.

In the bichromatic case, however, no simple expression for the cutoff energy
exists \cite{bichro,cfbichro1,cfbichro2,doublemil}, so that it is not
straightforward to predict whether the plateau can be extended. As a general
feature, high-harmonic spectra from atoms in bichromatic fields exhibit a
plateau with a relatively complex structure, which may have several cutoffs.
These cutoffs are given by local maxima of $E_{{\rm kin}}(t_{0},t_{1}),$
which are, however, unequally important for the resulting spectra. Thus, it
may happen that the absolute maximum of $E_{{\rm kin}}(t_{0},t_{1})$ leads
to a much less pronounced decrease in the harmonic signal than a local
maximum at a lower energy position.

A very illustrative example is a bichromatic field consisting of a wave with
frequency $\omega $ and its second harmonic \cite
{cfbichro1,cfbichro2,doublemil}. The addition of the second driving wave
causes a splitting in the monochromatic-field cutoff energy $\varepsilon
_{\max }=$ $|\varepsilon _{0}|+3.17U_{p},$ $U_{p}$ being the ponderomotive
energy. As a direct consequence, there is a double-plateau structure in the
harmonic spectra, with an upper and a lower cutoff, whose energies are
higher and lower than $\varepsilon _{\max }$, respectively. The upper cutoff
corresponds to the absolute maximum for $E_{{\rm kin}}(t_{0},t_{1}).$
However, it appears in the spectrum only as a small shoulder due to the
relative low harmonic intensity. The lower cutoff, on the other hand, is
related to a decrease of orders of magnitude in the harmonic yield, being
therefore the one effectively measured in experiments (see, e.g., \cite
{cfbichro2,andiel} for a more complete discussion). One can explain the
intensity difference in this double-plateau structure in terms of the width
of the effective potential barrier through which the electron tunnels. This
barrier is given by $V_{{\rm eff}}=V(x)-xE(t_{0})$, where $E(t_{0})$ is the
field at the electron emission time and $V(x)$ the atomic potential. For the
upper cutoff, the atomic potential is not as much distorted by the field as
for the lower cutoff. This results in a considerably wider effective
potential barrier and, consequently, much weaker harmonics.

In this paper we consider again the case of a linearly polarized $\omega
-2\omega $ field, as in \cite{cfbichro1,cfbichro2}. However, this time our
aim is to increase the intensities of the upper-cutoff-harmonics close to
the intensities of those belonging to the lower cutoff, effectively
extending the cutoff energy beyond $\varepsilon _{\max }=$ $|\varepsilon
_{0}|+3.17U_{p}$. For this purpose, one must provide an additional mechanism
for the electronic wave packet to reach the continuum, thus compensating the
weak tunneling ionization. We demonstrate that this can be done with an
additional driving wave. This third wave has a relative high frequency but
low intensity compared to the bichromatic field. The nature of the third
wave is such that it alters the electron injection into the continuum and,
at the same time, it does not appreciably modify the ponderomotive energy
and the cutoff. This field configuration is similar to that used in \cite
{xray}, where the influence of the background field on x-ray-atom scattering
processes was investigated.

We restrict ourselves to a one-dimensional model, which still describes
high-harmonic generation with linearly polarized fields well in qualitative
terms. We solve the time-dependent Schr\"{o}dinger equation 
\begin{equation}
i{\frac{d}{dt}}|\psi (t)\rangle =\left[ {\frac{p\sp{2}}{2}}+V(x)-p\cdot
A(t)\right] |\psi (t)\rangle 
\end{equation}
numerically, for an atom initially in the ground state, with binding
potential $V(x)$ subject to a laser field $E(t)=-dA(t)/dt$. Atomic units are
used throughout. The external laser field is taken as 
\begin{equation}
E(t)=E_{01}\sin (\omega _{1}t)+\sum_{i=2}^{3}E_{0i}\sin (\omega _{i}t+\phi
_{1i}),  \label{field}
\end{equation}
with $E_{0i},\ \omega _{i}$ and $\phi _{1i}$ being the field amplitudes,
frequencies and the phases with respect to the first driving wave,
respectively. In this paper, we choose $\omega _{1}=\omega ,$ $\omega
_{2}=2\omega ,$ $E_{03}\ll E_{01},E_{02}$ and $\omega _{3}\gg \omega .$
Unless stated otherwise, the relative phase $\phi _{13}$ is set to zero. The
binding potential is taken as 
\begin{equation}
V_{G}(x)=-\alpha \exp (-x^{2}/\beta ^{2}),  \label{potential}
\end{equation}
which is a widely used expression for modelling short-range potentials. The
harmonic spectra are calculated from the dipole acceleration $\ddot{x}%
=\left\langle \psi (t)\right| -dV(x)/dx+E(t)\left| \psi (t)\right\rangle $ 
\cite{dipacc}$.$ Spectral and time-frequency analysis is subsequently done
on  $\stackrel{..}{x}(t)$. This latter method has been extensively used to
extract the main contributions to high-harmonic generation within a cycle of
the driving field \cite{cfbichro1,wavelet}. These contributions give the electron
return times corresponding to a particular set of harmonics
(see \cite{cfbichro1,wavelet} for a more complete discussion). 

Time-resolved spectra are computed by performing a Fourier transform with a temporally
restricted window function. For an arbitrary function $f(t^{\prime })$, this
transform is

\begin{equation}
{\cal F}(t,\Omega ,\sigma )=\int\limits_{-\infty }^{+\infty }dt^{\prime
}f(t^{\prime })W(t,t^{\prime },\Omega ,\sigma )\quad .
\end{equation}
We consider a Gabor transform, which has a gaussian window function 
\begin{equation}
W(t,t^{\prime },\Omega ,\sigma )=\exp [-(t-t^{\prime })\sp 2/\sigma
^{2}]\;\exp [{\rm i}\Omega t^{\prime }]\ .
\end{equation}
The usual Fourier transform ${\cal F}(\Omega )$, for which all the temporal
information is lost, is recovered for $\sigma \rightarrow \infty $. The
temporal width $\sigma $ corresponds to a frequency bandwidth $\sigma
_{\Omega }=2/\sigma $.

The kinetic energy of the electron upon return is taken as 
\begin{equation}
E_{{\rm kin}}(t_{0},t_{1})=\frac{1}{2}\left[ A(t_{1})-A(t_{0})\right] ^{2}.
\label{kinetic}
\end{equation}
The ponderomotive energy is given by 
\begin{equation}
U_{p}=\frac{1}{2}\int_{0}^{T}A^{2}(t)dt=\sum_{i=1}^{3}\frac{E_{0i}^{2}}{%
4\omega _{i}^{2}}.  \label{pond}
\end{equation}
To first approximation, if the third driving wave is much weaker than the
others and its frequency is much higher than $\omega $, the contribution
from the additional field to the ponderomotive energy (\ref{pond}) and to
the kinetic energy (\ref{kinetic}) can be neglected. More specifically, for
the parameters used in this paper, we observed, after solving the classical
equations of motion of an electron in a field (\ref{field}), that for $%
E_{03}/E_{02}\leq 1$ and $\omega _{3}>5\omega $ the influence of the third
driving wave on these two latter quantities was not significant. 

As a starting point, we shall discuss the existence of the enhancement in
question. We will restrict ourselves to varying the relative phase $%
\phi_{12} $ between the two low-frequency driving waves and the
high-frequency field parameters. The bichromatic field strengths are similar
to these in \cite{cfbichro2}, namely $E_{01}=0.1\ {\rm a.u.}$, $%
E_{02}=0.032\ {\rm a.u},$ which give the intensity ratio $I_{2\omega
}/I_{\omega }=0.1.$ For these field parameters, the high-harmonic spectrum
displays a clear double-plateau structure, with a lower and an upper cutoff
(c.f. Fig.~1 and Refs. \cite{cfbichro1,doublemil}). The lowest frequency is taken as $\omega =0.057\ {\rm a.u.},$
which is typically used in experiments. The ground-state energy was taken as 
$|\varepsilon _{0}|=0.57\ {\rm a.u.},$ which roughly corresponds to the
argon ionization potential and, unless stated otherwise, we took $%
\alpha=1.15 \ {\rm a.u.}$ and $\beta=1\ {\rm a.u.}$ in (\ref{potential}).
This gives a model-atom with a single bound state.

The influence of the third driving wave on the harmonic spectra is shown in
Fig.~1, for relative phases $\phi _{12}=0$ and $\phi _{12}=0.3\pi $ and
several field strengths $E_{03}.$ The relative phase $\phi _{13}$ was set to
zero in both cases, and $\omega _{3}$ was chosen to be ten times $\omega_1$.
Apart from the strong enhancement around $\omega _{3}$ (near harmonic order
10), the main differences between the case with a third wave and the
bichromatic case $(E_{03}=0)$ are displayed in the so-called
``upper-cutoff'' harmonics, which lie, for the phases in question, roughly
at $\varepsilon_u=|\varepsilon_0|+3.8U_p$, with the corresponding harmonic
orders near $N=62$. The intensity of this group of harmonics can be changed
significantly when the field strength of the third wave is increased.
Whereas for $\phi_{12}=0$ these changes are quite irregular, for $\phi
_{12}=0.3\pi $ they appear as very pronounced enhancements, which reach
three orders of magnitude. They make the intensities of the upper-cutoff and
lower-cutoff harmonics comparable, effectively extending the harmonic
production region. This occurs already for a third wave whose intensity is
only a few percent of that of the bichromatic field.
\begin{figure}[tbp]
\epsfig{file=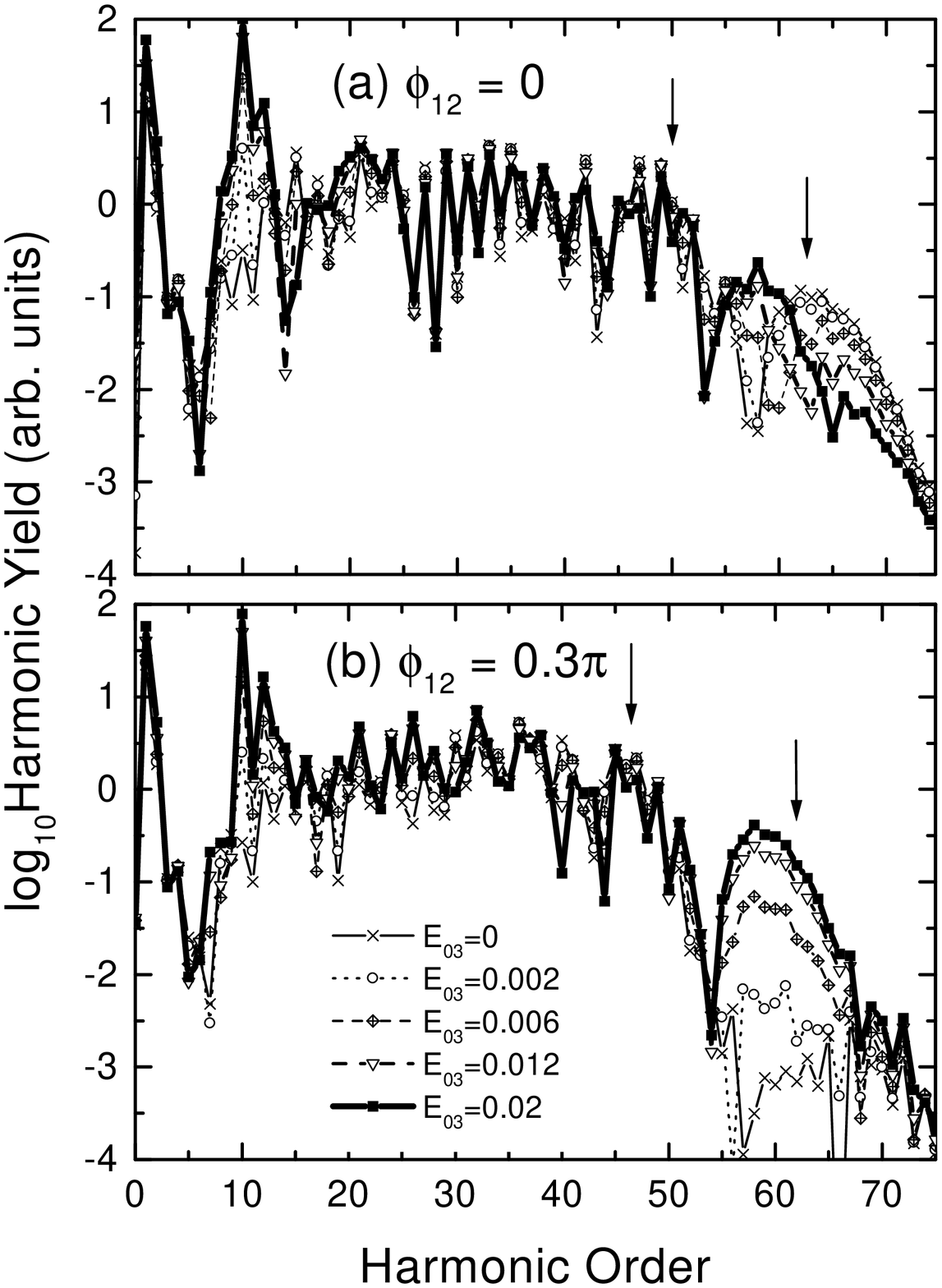,width=7.0cm,angle=0}
\caption{Harmonic spectra for an atom subject to the three-color
field ({\protect\ref{field}}), with strong-field amplitudes $E_{01}=0.1 \ {\rm a.u.}$, 
$E_{02}=0.032 \ {\rm a.u.}$, high-frequency field strengths $0\leq E_{03}\leq 0.02  \ {\rm a.u.}$,
 frequencies $\omega_1=0.057\ {\rm a.u.}$,  $\omega_2=0.114\ {\rm a.u.}$
and  $\omega_3=0.57\ {\rm a.u.}$, and relative phase $\phi _{13}=0$,
for $\phi _{12}=0$ (part (a)) and  $\phi _{12}=0.3\pi$ (part (b)).
The lower cutoff (near harmonic order 50) and the upper cutoff (near 
harmonic order 62) are marked with arrows. The lower-cutoff energy depends more sensitively on $\phi_{12}$ than the upper-cutoff
energy [5]. }
\end{figure}

A particularly interesting feature of the scheme is that it affects mainly
the group of harmonics in the upper cutoff, leaving other groups of
harmonics practically unaffected. This is in contrast with the purely
bichromatic case, for which a change in the relative phase $\phi _{12}$
leads to changes in the whole plateau structure \cite{cfbichro1,cfbichro2}.
In analyzing the enhancement effects, we recall that the presence of the
high-frequency third wave does not modify the ponderomotive energy and the
classical trajectories. Thus, the three-step model leads us to the
conclusion that a high-frequency induced process is injecting the electron
into the continuum. This physical picture also explains why the remaining
harmonic intensities are not influenced by the high-frequency wave. Let us
consider an
electron leaving the atom at the emission time $t_{0l}$, so that, 
at its recombination time $t_{1l}$, the lower-cutoff harmonics are
 generated. At the lower-cutoff emission time $t_{0l}$, tunneling is 
 so pronounced that, even in the presence of the
third driving wave, this mechanism is still the dominant path for the 
electron to reach
the continuum. Thus, the high-frequency induced process is in comparison
negligible. A similar argument explains the differences between Figs.~1(a)
and 1(b). For $\phi _{12}=0$, there is still enough tunneling to compete
with the high-frequency induced process, such that the quantum interference
between both processes leads to irregular intensity variations. For $\phi
_{12}=0.3\pi $, on the other hand, tunneling is strongly suppressed, such
that the electron reaches the continuum mainly due to the high-frequency
induced process \cite{cfbichro2}.
\vspace{-0.5cm}
\begin{figure}[tbp]
\epsfig{file=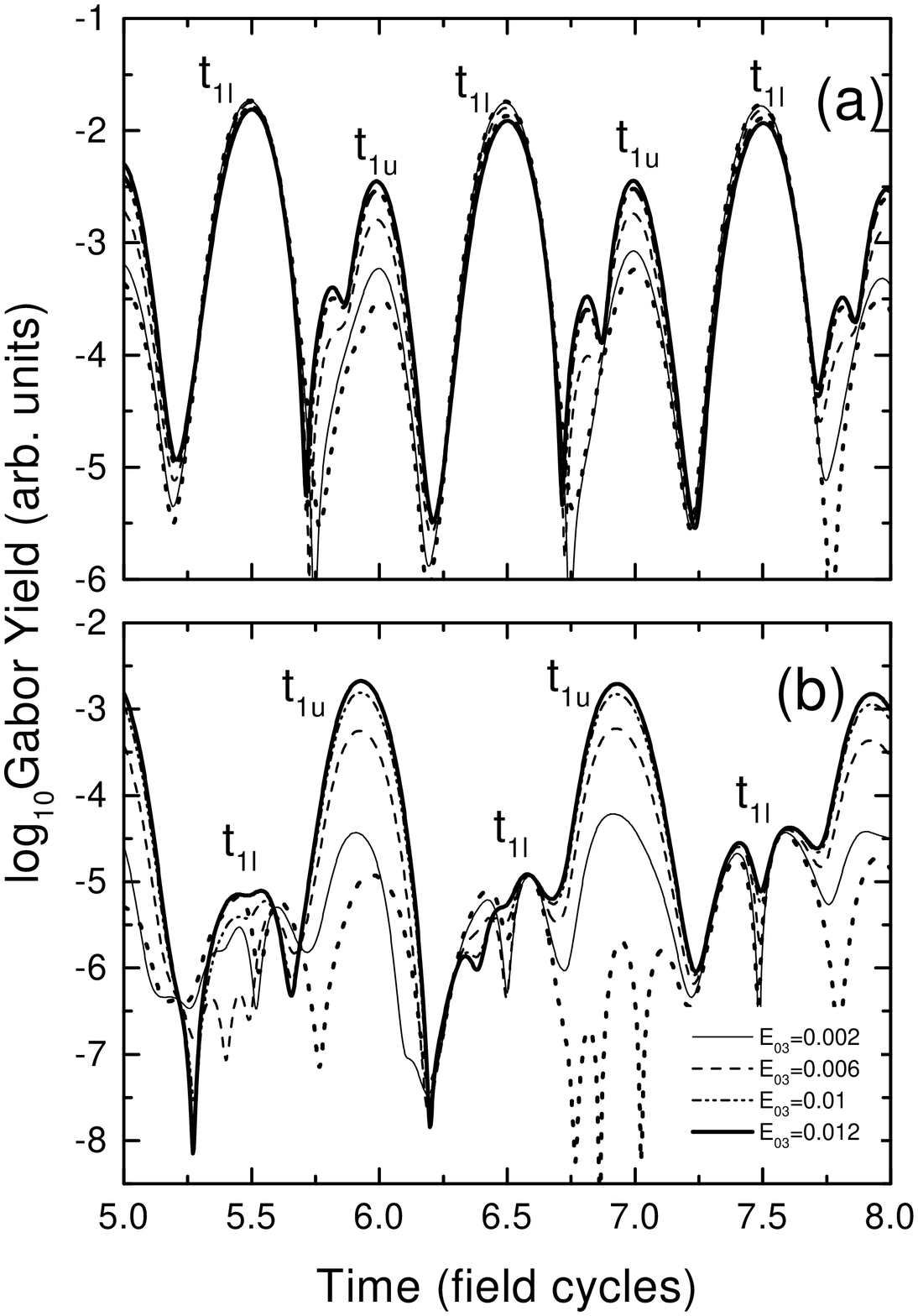,width=7cm,angle=0}
\caption{Time-resolved spectra for the lower (part (a)) and upper 
(part (b))cutoff 
harmonics, for the same bichromatic-field parameters as in Fig.~1, $\phi _{12}=0.3\pi$ and  $\phi _{13}=0$, and several 
high-frequency field amplitudes $E_{03}$. The thick dotted lines correspond
to $E_{03}=0$. The temporal width was
chosen as $\sigma=0.1T$, such that the frequency width $\sigma_{\omega}$
is of roughly three
harmonics. The precise center of the window function has been chosen as
$\Omega_l=48\omega$ (part (a)) and $\Omega_u=58\omega$ (part (b)). The return
times $t_{1l}$ and $t_{1u}$ are indicated in the figure. }  
\end{figure}

This interpretation is supported by the time-frequency analysis. In Fig.~2,
we display the time-frequency yield for groups of harmonics centered at
the frequencies $\Omega _{l}=48\omega$ (Fig.~2(a)) and $\Omega _{u}=58\omega$ 
(Fig.~2(b)), 
for several field strengths $E_{03}$ and a relative
phase $\phi _{12}=0.3\pi $. The frequency $\Omega_l$ corresponds to the
lower-cutoff energy and $\Omega_u$ was chosen at the most prominent harmonic frequency in the upper cutoff. These time-resolved spectra are periodic within a
cycle $T=2\pi/\omega$ of the driving field, exhibiting peaks at times
$t_{1l}=1.4T$ and $t_{1u}=0.9T$. These are the electron return times
 related to the lower- and the upper-cutoff harmonics,
respectively, and have been computed in \cite{cfbichro2}, together with the
corresponding emission times (c.f. Fig.~1 in there). 
For a given set of harmonics, if  the high-frequency
field is enhancing the injection of the electron wave packet at the emission
time $t_{0}$, the corresponding peak at 
$t_{1}$ in the time-resolved spectra should get more prominent 
as $E_{03}$ is increased. This occurs only for the upper-cutoff return time
$t_{1u}$. This can be observed in both parts of the figure.

In the following we understand how the observed enhancements depend on the
relative phase $\phi _{13}$ between the lowest- and highest-frequency
fields, on the field strength $E_{03}$, on the frequency $\omega _{3}$, and
whether the atomic potential has any influence on it.

The behavior with respect to $\phi _{13}$ gives further information about
the nature of the effect due to the third wave. In case this wave enhances
tunneling ionization by changing the effective potential barrier, or
increases the harmonic intensities by distorting the electron motion in the
continuum, this influence will be different for different phases $\phi _{13}$. 
We verified that the upper-cutoff harmonics remain unchanged
when $\phi _{13}$ is varied. 
This result suggests that the third wave is playing a
very different role than the bichromatic field. These results are 
displayed in Fig.~3.
\begin{figure}[tbp]
\epsfig{file=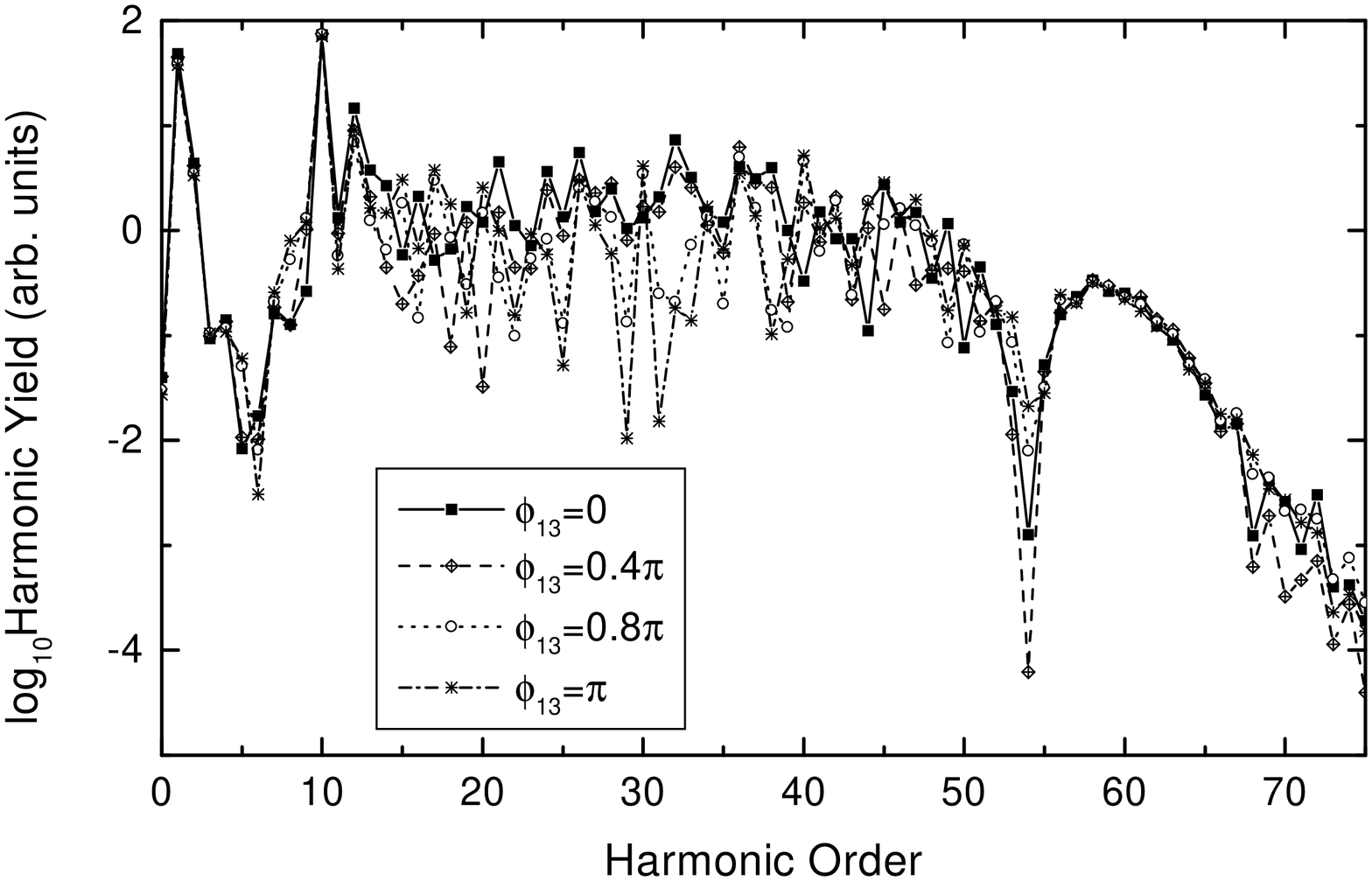,width=5.5cm,angle=270}
\caption{Harmonic spectra for the same bichromatic-field parameters as in the
previous figures, $\phi _{12}=0.3\pi $, field strength $E_{03}=0.015$ a.u., 
and several relative phases $\phi _{13}$.}  
\end{figure}
Another very important issue is the dependence of the enhancements on the
field strength $E_{03}$. A closer inspection of Figs.~1(b) and 2(b) suggests
that there is a saturation intensity for the upper-cutoff harmonics (in the
figure, close to $E_{03}=0.01\ {\rm a.u.}$). Furthermore, these harmonics
are unequally enhanced, the maximum enhancement occurring, for the example
in question, at the harmonic order $N=58$. This maximum can be slightly
displaced, depending on the frequency $\omega _{3}$. In Fig.~4, we show
explicitly the behavior of this maximally enhanced harmonic with respect to
the field strength $E_{03}$, for several frequencies $\omega _{3}$. This
figure shows that the enhancement effect increases quickly in the weak field
region but it eventually saturates when reaching the strong field region,
which means that upper-cutoff harmonics can not be indefinitely enhanced by
increasing the strength of the high frequency wave. 
\vspace{-0.5cm}
\begin{figure}[tbp]
\epsfig{file=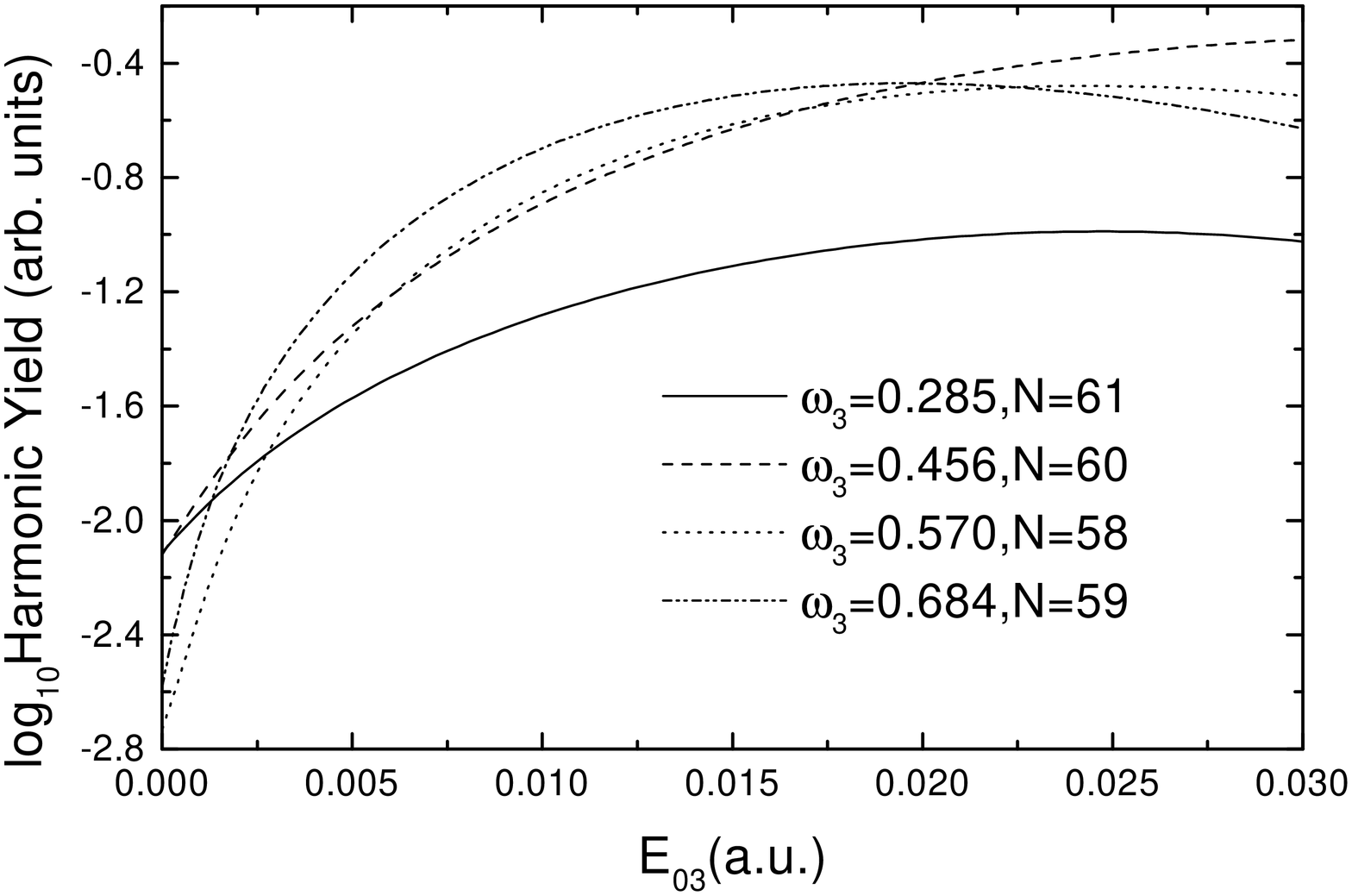,width=5.5cm,angle=270}
\caption{Harmonic yields as functions of the high-frequency field strength 
$E_{03}$, for the same field as in the previous figures,
 with $\phi _{12}=0.3\pi$ and
 $0.285  \ {\rm a.u.}\leq \omega_3 \leq 0.684  \ {\rm a.u.}$.
 The harmonic orders displayed in the figure correspond to the most efficient
enhancements obtained.}  
\end{figure}
\begin{figure}[tbp]
\epsfig{file=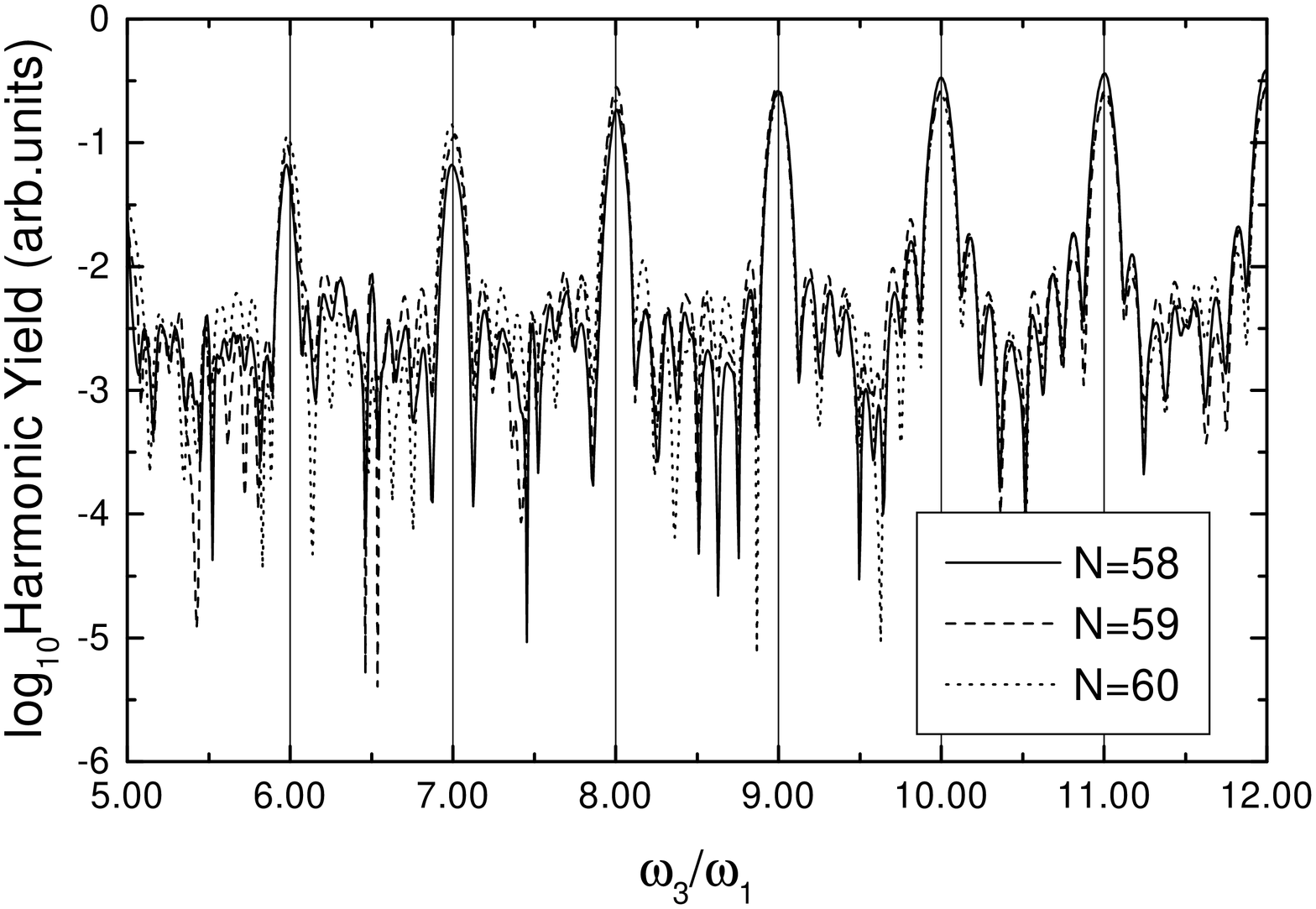,width=6cm,angle=270}
\caption{Harmonic yields of neighboring upper-cutoff harmonics
 as functions of the frequency ratio
$\omega_3/\omega_1$, for a driving field as in the previous figures and
$E_{03}=0.015 \ \mathrm{a.u.}$. The ionization threshold is at
$\omega_3=10\omega_1$. }
\end{figure}
In Fig.~5 we show the dependence of the enhanced harmonics with respect to
the frequency $\omega _{3}$, for $5\omega _{1}\leq \omega _{3}\leq 12\omega
_{1}$. The figure displays major peaks at integer frequency ratios $\omega
_{3}/\omega _{1}$ and sub-peaks between the integers. The enhancements are
specially strong at the integers, where the frequencies of the three driving
waves are commensurate. The strong enhancements at these particular
frequencies can be understood as a consequence of the high-harmonic
generation process: since the production of harmonics takes place within
many cycles of the driving field, one expects it to be most efficient when
this field is periodic \cite{bremstrahl}. Fig.~5 also demonstrates that the
observed enhancements are not related to threshold effects, since they occur
when the frequency of the third wave is either below or above the threshold.

The investigations presented above are from our numerical calculations for
the short-range potential (\ref{potential}), with $\alpha $ and $\beta $
chosen in such a way that it has a single bound state. To rule out the
possibility that the effect is due to an artifact introduced by this model
potential, we performed similar calculations for several potentials with
different values for depth $\alpha $ and width $\beta $. We also studied the
soft-core potential $V_{C}=-\alpha \left[ x^{2}/\beta ^{2}+1\right] ^{-1/2}$%
. We observed that the enhancements are present in all situations, being
however stronger for short-range potentials. This is related to the fact
that tunneling ionization is more efficiently suppressed in the short-range
case, such that it does not compete with the high-frequency induced process.
For the gaussian potential used, we obtain very pronounced enhancements for $%
\beta \leq 2\ {\rm a.u.}$, which is well within the experimental range.

In conclusion, we have shown that, with an additional high-frequency field,
we are able to enhance a group of very high harmonics of a bichromatic
driving field consisting of a wave of frequency $\omega $ and its second
harmonic. These harmonics have energies up to $|\varepsilon _{0}|+4U_{p}$,
which is about 30\% higher than the monochromatic cutoff frequency $%
\varepsilon _{{\rm max}}=|\varepsilon _{0}|+3.17U_{p}$. They correspond to a
set of electron trajectories for which tunneling is not very pronounced in
the bichromatic case. The high-frequency field provides an additional
mechanism for the electron to reach the continuum, resulting in the
enhancement of this group of harmonics.

This physical interpretation is supported by several 
characteristics of the enhancements, such as, for instance, the fact that 
the remaining groups of harmonics 
are not altered by the high-frequency field. 
For an electron leaving at a time $t_0$ such that the lower-cutoff and
plateau harmonics are generated, tunneling ionization is the dominant process.
This is true even in the presence of 
the third wave, so that the high-frequency induced process 
plays practically no role. For the upper-cutoff harmonics, on the other hand,
tunneling is relatively weak, so that the high-frequency induced process may
compete with it or even be far more prominent. This latter effect is 
observed in the behavior of the enhancements with respect to the 
relative phase $\phi _{12}$ between
the two low-frequency waves.
 The time-resolved spectra of the lower -and upper cutoff harmonics confirm
this physical picture.  As the high-frequency field is increased, the
time-frequency yield at the return times $t_{1l}$ corresponding to the 
lower cutoff remains
pratically unaltered, whereas the yield at the upper-cutoff return time
$t_{1u}$ is enhanced in orders of magnitude. Thus, the third field is
increasing the electron injection in the continuum at the corresponding 
electron emission times $t_{0u}$. Finally, since 
the enhancements are independent of the phase 
$\phi_{13}$ between the highest- and lowest frequency fields, they can not 
be atributted to mechanisms such as tunneling ionization or a distortion in
the electron propagation. 

The enhancements are particularly strong when the relative phase between 
the two strong driving waves is chosen such that tunneling ionization 
is strongly suppressed. Such a case is provided by taking 
$\phi _{12}=0.3\pi $ \cite{cfbichro2}, for which an appreciable enhancement
is already obtained with high-frequency fields whose intensities are only a 
few percent of that of the bichromatic fields. This phase control is 
already experimentally
feasible \cite{andiel}. Another prerequisite for pronounced enhancement
effects concerns the frequency of the third wave, which must be an integer
multiple of the fundamental frequency of the bichromatic field. Furthermore,
our theoretical studies show that the enhancements are always present for
very different potentials, but they also suggest that a short-range
potential gives stronger enhancements and therefore is a more appropriate
choice for a possible experimental consideration.

It is also worth mentioning that, in principle, a similar enhancement scheme
is applicable to any sets of harmonics corresponding to a set of electron
trajectories for which tunneling ionization is strongly suppressed. For the
specific $\omega -2\omega $ configuration considered in this paper, this
condition is satisfied by the upper-cutoff harmonics.

Acknowledgments: We are grateful to A. Fring and to J.M. Rost for useful
discussions. 
\end{document}